\documentstyle[12pt]{article}
\begin{document}
\setlength{\baselineskip}{24pt}
%Fin Version preprint

\newcommand{\fig}[3]{
\begin{figure}
\vspace{#2}
\label{#1}
\caption{#3}
\end{figure}
}
\newcommand{\hs}{\hspace{1cm}}
\newcommand{\be}[1]{\begin{equation}\label{#1}}
\newcommand{\BE}{\begin{equation}}
\newcommand{\ee}{\end{equation}}
\newcommand{\bee}{\begin{eqnarray}}
\newcommand{\eee}{\end{eqnarray}}

\newcommand{\Npp}{N_{++}}
\newcommand{\Np}{N_+}
\newcommand{\Nm}{N_-}

\newcommand{\Cp}{\chi_+}
\newcommand{\Cm}{\chi_-}

\title{Hysteretic behavior
of a polymer molecule immersed in an incompatible melt}
\author{Cyprien Gay (i), Elie Rapha\"el (ii),\\
(i) Laboratoire CNRS - Elf Atochem (UMR 167),\\
95, rue Danton, B.P.108, 92303 Levallois-Perret cedex, France,\\
(ii) Coll\`ege de France,
Physique de la Mati\`ere Condens\'ee \\
(URA 792 du CNRS),\\
11, place Marcelin Berthelot,\\
 75231 Paris cedex 05, France.}
\date{\today}
\maketitle

\begin{abstract}
The radius of gyration of a polymer chain
immersed in a low molecular weight solvent
is known to vary monotonically
with the solvent quality.
Here, we consider the behavior of a chain immersed
in a high molecular weight solvent
(polymer melt).
Unsurprisingly, we find that,
as the incompatibility between the chain
and the polymer melt is increased,
the two limiting conformations
of the chain
are the ideal random coil ($R\simeq aN^{1/2}$)
and the dense globule ($R\simeq aN^{1/3}$).
We show, however, that between these
two limits the radius of gyration of the chain
presents a hysteretic behavior .

\end{abstract}

%%%%%%%%%%%%%%%%%%%%%%%
\section*{Introduction}
%%%%%%%%%%%%%%%%%%%%%%%

The behaviour of an isolated linear chain
(with degree of polymerization N)
dissolved in a solvent of low molecular weight is well understood
\cite{1,2,3}.
The effective interaction between two monomers
may be described in terms of the Flory
parameter $\chi$, which, in most cases, is positive.
For $0 \leq \chi \leq \frac{1}{2} (1 - \frac{1}{\sqrt{N}})$,
the chain is swollen and its radius, $R$, is given by
$R \simeq a N^{3/5} (1 - 2 \chi)^{1/5}$
where $a^3$ is the monomer volume.
For $\frac{1}{2} (1 - \frac{1}{\sqrt{N}})
\leq \chi \leq \frac{1}{2} (1 +\frac{1}{\sqrt{N}})$,
the chain can be described as a random walk and $R \simeq a N^{1/2}$.
For $\frac{1}{2} (1 + \frac{1}{\sqrt{N}}) \leq \chi \leq 1$,
the chain is collapsed
and it radius is given by $R \simeq a N^{1/3} (- v/a^3)^{2/3}$.
For $\chi$ larger than unity, $R$ is given by $R \simeq a N^{1/3}$.
The radius of the chain therefore decreases monotonically
when $\chi$ increases, that is when the quality of the solvent
(described by the excluded volume parameter
$v = a^3 (1 - 2 \chi)$) decreases,
and no hysteresis in the chain conformation
is expected as  $\chi$ is varied \cite{WUWANG}.

Consider now an isolated linear chain, with degree of polymerization  $N$,
dissolved in a melt of {\it chemically identical} chains with degree of
polymerization $P$.
This situation is also well understood \cite{1,2,3}.
The bare monomer-monomer interaction of the $N$-chain, described by the
excluded-volume parameter $a^3$, are screened out by the $P$-chains,
and the effective excluded volume parameter is given by $a^3/P$
\cite{4,5}.
Consequently, the $N$-chain
remains ideal (i.e. $R \simeq a N^{1/2}$) as long as $N \leq P^2$.
However, for $N >P^2$,
the chain is swollen
and its radius is given by
$R \simeq a N^{3/5}P^{-1/5}$ \cite{6,7,8,9}.
In the present
article, we generalize this analysis to the case
where the $N$-chain and the $P$-chains
are {\it chemically different} (see the work
of Joanny and Brochard {\cite{JFFB}}).
Consider for instance the limit $P\rightarrow \infty$.
For $\chi = 0$, the chain radius is given by $R \simeq a N^{1/2}$
while for large values of $\chi$ (i.e. $\chi \gg 1$),
we expect $R$ to be given by  $R\simeq a N^{1/3}$.
We here investigate the whole range of incompatibilities,
both for $P\rightarrow \infty$ and for finite $P$.

%%%%%%%%%%%%%%%%%%%%%%%
\section*{Free energy}

Consider a chain ($N$ monomers of size $a$)
immersed in a melt of chemically different chains
($P$ monomers, Flory interaction $\chi$
between both types of monomers).
The Flory free energy \cite{4} of the system
can be expressed as:
\be{FLORY}
\frac{F}{kT}=\frac{R^2}{Na^2}+\frac{Na^2}{R^2}
+N\chi\left(1-\frac{Na^3}{R^3}\right)
+\frac{N^2 a^3}{PR^3}
+\frac{N^3a^3}{PR^6}
\ee
The first and second terms
are the classical elastic and confinement terms.
The third term corresponds to enthalpic interactions between the $N$-chain
and the $P$ chains:
for each monomer of the $N$-chain,
these interactions amount to $\chi kT$
per contact with a melt monomer
(probability $1-\phi_N$, where $\phi_N=Na^3/R^3$
is the volume fraction of the $N$-chain).
The last two terms describe the screened
excluded volume interactions \cite{5} (see Introduction).
The radius of the chain is obtained by minimizing  the free energy,
along with the condition  $R\geq aN^{1/3}$.
For the moment, we consider
the limit of high melt molecular weights
($P\rightarrow\infty$) and therefore neglect the last two
terms on the {\it{r.h.s.}} of eq. (\ref{FLORY}).

The general shape of the above free energy,
as a function of the radius of gyration $R$,
depends on the degree of incompatibility:
three different regimes arise, depending on the value
of the Flory interaction parameter $\chi$
compared to the following values:
\bee
\Cm&\equiv&N^{-2/3}\label{Cm}\\
\Cp&\equiv&N^{-1/2}\label{Cp}
\eee
In the intermediate regime $\Cm<\chi<\Cp$,
the free energy has two different minima (see Fig~1):
one that corresponds to a Gaussian chain
($G$, radius of gyration $aN^{1/2}$),
and one that corresponds to a dense globule ($D$),
whose radius of gyration is very close to $aN^{1/3}$.
In the weak and strong incompatibility regimes, conversely,
there is only one minimum:
the chain is unperturbed ($G$) if the incompatibility
is weak ($\chi<\Cm$),
and it is collapsed into a dense globule ($D$)
in the reverse case ($\chi>\Cp$).

We now focus on the intermediate regime ($\Cm<\chi<\Cp$),
where two conformations are possible,
and we discuss their relative stabilities.
The more stable conformation is the dense globule $D$
(except when $\chi$ approaches $\Cm$).
However, since $\chi<\Cp$, the amplitude of the barrier
$\Delta F\simeq kT/(N\chi^2)$,
is large compared to thermal fluctuations
and the chain retains a Gaussian conformation.
It is in fact necessary to go into some more detail
in order to fully justify the above statement.
Indeed, the above Flory approach is global (mean field).
Hence, the sole consideration of equation (\ref{FLORY})
only shows that the monomers of the chain
have no tendency
to collapse in a collective manner.
But one can wonder whether the chain
could tend to collapse locally,
which would progressively lead
to the dense collapsed state $D$,
possibly without any need for passing high energy barriers.
One can easily show that this is  not the case.
Indeed, the barrier $kT/(m\chi^2)$
for a subchain containing $m$ monomers ($m<N$)
is even higher than the barrier for the whole chain:
there is no tendency
towards any local collapse of the chain.
The global, Flory approach is thus sufficient.

Since the amplitude of the barrier
in the intermediate regime is high
(except near $\Cm$ or $\Cp$),
the radius of gyration of the chain
presents an hysteretic behavior
as the incompatibility $\chi$ is varied.
If we start from a low value of $\chi$ ($\chi<\Cm$)
and progressively increase  $\chi$,
the chain remains Gaussian ($R\simeq aN^{1/2}$)
as long as $\chi < \Cp$.
For $\chi > \Cp$, the chain collapses and
reaches its equilibrium conformation,
namely that of a dense globule ($R\simeq aN^{1/3}$).
If, on the other hand, we start from a high value of
$\chi$ ($\chi>\Cp$) and progressively decrease  $\chi$,
the chain remains collapsed as long as $\chi > \Cm$
and recovers its Gaussian conformation only for $\chi<\Cm$.
The fact that both antagonistic transitions
(collapse and swelling) take place
for different degrees of incompatibility ($\Cp\neq\Cm$)
constitutes a hysteretic behavior.

%%%%%%%%%%%%%%%%%%%%%%%
\section*{Interpretation of the hysteresis}

The above Flory approach describes
the hysteretic behaviour of the system.
To understand the physical origin of the hysteresis,
we present below two alternative interpretations.
The first one is an analogy with adsorption
(the collapse of the $N$-chain is analog to self-adsorption).
The second interpretation is based on the fact
that the melt chains may or may not penetrate
the volume pervaded by the $N$-chain,
depending on the degree of incompatibility.

Let us start with the first interpetation.
In the classical scaling theory of polymer adsorption \cite{2},
each adsorbed blob ($g$ monomers)
gains an enthalpy $n\epsilon$ through the contact
of $n$ of its monomers with the surface
(typically $n=g^{1/2}$ for a Gaussian blob).
This energy gain is balanced by a free energy loss
(confinement entropy) of the order of $kT$ per blob.
Hence, the number of monomers per blob
is given by $n\epsilon=kT$, which reads:
$g=(kT/\epsilon)^2$.
The gain in free energy that results from this balance
is of order $kT$ per blob.
Therefore, for the chain to adsorb,
it has to contain at least one blob, {\it i.e.}
$N\geq(kT/\epsilon)^2$.

Consider now the present situation of one $N$-chain
immersed in an incompatible melt.
The equilibrium conformation
of a confined chain ($R\leq aN^{1/2}$)
can be described in terms of $N/g$ superimposed blobs,
each containing $g$ monomers (with $R=ag^{1/2}$).
Typically, the number of contacts
between one blob and other monomers of the chain
is $n=g\phi$, where $\phi=Na^3/R^3$
is the overall chain volume fraction.
If the confinement of the chain
results from the incompatibility with the $P$ melt,
its equilibrium conformation can be seen
in terms of self-adsorption:
if the incompatibility (measured by $\chi$)
is strong enough
and if the volume fraction is sufficiently high,
the contacts between monomers of the chain
provide a sufficient enthalpic gain
to yield a certain blob dimension
(i.e., radius of gyration).
Whereas if $\chi$ is small,
the corresponding blob dimension is too large ($g>N$),
and the chain is then Gaussian.
Following the analogy with the problem
of an adsorbed chain on a surface and
noticing that here, $\epsilon=\chi kT$,
the equilibrium between enthalpy gain
and confinement entropy
can be written as
$n\chi kT=kT$, i.e.,
\be{aNchi}
g\phi\chi=1
\ee
The fact that the above criterion
depends on the chain volume fraction $\phi$
is the very reason for the existence
of the hysteresis.
Indeed, it causes the corresponding adsorption thresholds
for the Gaussian conformation and for the dense globule
to be different.
For the Gaussian chain, for instance,
the blob is the whole chain ($g=N$)
and the volume fraction is small ($\phi=N^{-1/2}$).
The corresponding incompatibility threshold
is $\chi=\Cp=N^{-1/2}$.
Hence, as long as $\chi<\Cp$,
the chain is not self-adsorbed (Gaussian conformation).
Once it reaches $\Cp$, it starts to self-adsorb.
The volume fraction correspondingly increases,
which in turn reinforces the tendency
towards self-adsorption.
This process only stops when steric hindrances
prevent the chain any from further collapse,
namely when $R\simeq aN^{1/3}$.

Conversely, when the chain is a collapsed globule,
its volume fraction $\phi$ is equal to unity,
and the blob corresponds to $g=R^2/a^2=N^{2/3}$ monomers.
The corresponding threshold is $\chi=\Cm=N^{-2/3}$
(which is lower than the Gaussian threshold $\Cp$).
Hence, if $\chi$ is above $\Cm$,
the globule is stable.
If $\chi$ becomes lower than $\Cm$, however,
the chain cannot continue to remain self-adsorbed.
It therefore starts to swell.
This reduces its volume fraction,
which in turn enhances its swelling,
until the chain reaches its equilibrium,
ideal, Gaussian conformation.

An alternative derivation of equation (\ref{aNchi})
is based on the behaviour of the melt chains.
If the incompatibility is strong,
entering the volume pervaded by the $N$-chain
is too unfavourable for the $P$ chains,
and they rather go back into a pure melt region.
This can be described more quantitatively as follows.
For a $P$ chain, assuming that it remains Gaussian,
penetrating the $N$ region of extension $R$
takes up roughly $p=(R^2/a^2)=g$ monomers,
among which a fraction $\phi$
are in contact with the $N$-chain
(where $\phi\simeq Na^3/R^3$ is still
the volume fraction of the $N$-chain).
If the corresponding enthalpy $p\phi\chi kT$
is higher than $kT$,
then the $P$ chain rather goes back into the melt,
thus suffering an entropy loss of order $kT$.
This criterion again yields equation (\ref{aNchi}).
Note that in the case of the globular conformation ($\phi=1$),
where the melt chains rather turn back into the melt
after $p$ monomers have penetrated the $N$ region ($p<g=R^2/a^2$),
it is possible to estimate the width $\lambda\simeq ap^{1/2}$
of the interface between the globule and the melt
(see references \cite{10,11}
for the similar problem of the interface between two melts).
The interface width increases
when the incompatibility $\chi$ is reduced.
One can check that the transition from the globule
to the Gaussian conformation ($\chi\simeq\Cm$)
corresponds to the point where the interface
has invaded the whole globule ($\lambda\simeq R$).

We have shown that the consideration of equation (\ref{aNchi})
obtained through the analogy with adsorption
or alternatively through the consideration
of the melt chains,
proves that the chain conformation
has a hysteretic behaviour
when the Flory parameter $\chi$ is tuned.
It thus provides a simple physical interpretation
for the hysteresis obtained initially
through the Flory approach.

%%%%%%%%%%%%%%%%%%%%%%%
\section*{Shorter melt chains}

The above considerations show
that the radius of gyration of the $N$-chain
has a hysteretic dependence
on the Flory parameter $\chi$
in the limit of very long melt chains
($P\rightarrow\infty$).
In the opposite limit (low molecular weight solvent
with $P = 1$),
the radius of gyration is known
to be a single-valued function of $\chi$.
In the following,
we consider a melt with finite chain length $P$
in order to investigate the crossover
between hysteretic and non-hysteretic behaviours.

The essential point is that
when the melt chain length is decreased,
the translational entropy of the molecules
becomes comparatively more important.
When $P<\chi^{-1}$, the two-body term
in the Flory free energy (equation \ref{FLORY})
exceeds the enthalpic term.
As a result, the free energy has now only one minimum,
and the variations of the radius of gyration
with the degree of incompatibility
are then similar to the case
of a low molecular weight solvent.
The corresponding phase diagram
for the $N$-chain for all values of $P$
is depicted on Fig.~3.
In the swollen regime (S),
the radius of gyration is given by
$R\simeq aN^{3/5}P^{-1/2}(1 - 2 \chi)^{1/5}$ \cite{6}.
For $P\simeq 1$, we recover the known
bad-to-good solvent transition
which includes the $\theta$-solvent regime
for $\chi$ of order unity
($\chi=1/2$ in the classical description by Flory
recalled in the Introduction).
When $P\leq N^{1/2}$, as was shown by J.-F.~Joanny and F.~Brochard~\cite{JFFB},
a hysteresis is present between $\chi=(P^{-1}-N^{-1/2})/2$
and $\chi=(P^{-1}-N^{-1/2}P^{-1/4})/2$
(this zone appears only as a double line in the Log-Log diagram of Figure~2).
For $N^{1/2}\leq P\leq N^{2/3}$,
the width of the hysteresis range for $\chi$
increases with the melt molecular weight $P$
(since $\Cm\simeq P^{-1}$ and $\Cp={\rm const}$ in that regime).
For $P\geq N^{2/3}$,
the width is constant (equations \ref{Cm} and \ref{Cp}).

%%%%%%%%%%%%%%%%%%%%%%%
\section*{Conclusion}

A polymer chain immersed in a melt of long,
chemically different molecules,
adopts either a Gaussian conformation,
or a dense, collapsed conformation.
The transition between both situations
was shown to be hysteretic
when the degree of incompatibility is tuned
(this can be achieved for instance by varying the system temperature).
In practice, this effect will be more easily observed
at an interface, with scarcely grafted $N$-chains.
For such a polymer brush, the hysteretic behaviour
should combine with the tendency
to in-plane micellization \cite{12}
to yield a rich temperature dependence.
The transition could be used for tuning the interpenetration
between a brush and a matrix.
The hysteresis would then prove very useful
for designing a system that could be insensitive
to weak temperature fluctuations,
while it could be switched between both states
through larger temperature changes.
In order to obtain a good coupling
between the interfacial chains and the matrix,
large molecular weights are usually required.
This corresponds to rather small values of $\Cm$ and $\Cp$.
Such values can be obtained
for instance by using protonated $P$-chains
and deuterated $N$-chains (or vice-versa) \cite{13}.
Another possibility would be to use
a homopolymer melt ($P$-chains)
and an $N$-chain made of mainly the same monomer species
as the $P$-chains, copolymerized
with a very small fraction of chemically different monomers
(the number of comonomers and their incompatibility
must be small, in order to avoid microphase separation).
Kinetical aspects of the transition at $\chi\approx\Cm$
and $\chi\approx\Cp$ would require a separate development.
In the case of a simple solvent, many studies have been
devoted to this subject \cite{PGGCOLL,GROSBERG}.
Various intrachain effects may take place,
such as a glass transition or physical crosslinking in the globule, or
such as the formation of tight knots \cite{KNOT}
through sudden tension in a swelling entangled chain.
In a hypothetical system where none of these effects would
be dominant, we expect the kinetics to be governed by diffusion
of the molecules :
in the present case of a high molecular weight solvent,
the typical time-scale would therefore be the
therefore the reptation time
of the shorter chains ($N$ or $P$).\newline

{\large {\bf Acknowledgments}}. We thank P.-G de Gennes and T. Waigh for
useful discussions, and Ludwik Leibler for indicating reference~\cite{JFFB}
to us.

%%%%%%%%%%%% BIBLIOGRAPHIE automatique %%%%%%%%%%%%%%%%
%\bibliographystyle{BblED97}
%\bibliography{BIBSPC,BIB200,BIB400,BIB528,BIB560,BIB600}

\begin{thebibliography}{10}

\bibitem{1}
{P. J.} Flory,
\newblock {Principles of Polymer Chemistry}, Cornell University Press, NY, 1971.

\bibitem{2}
{P.-G.} de~Gennes, {\em Scaling Concepts in Polymer Physics},
\newblock Cornell Univ. Press, Ithaca, 1979.

\bibitem{3}
M.~Doi et {S. F.} Edwards, {\em The Theory of Polymer Dynamics},
\newblock Oxford Univ. Press, New York, 1986.

\bibitem{WUWANG}
Recent experiments, however, demonstrate a slight hysteresis
in very dilute water-solutions
of poly(N-isopropylacrylamide):
{C.} Wu and {X.} Wang,
\newblock {\em Phys. Rev. Lett.}, {\bf 80 (18)}:4092, 1998.

\bibitem{4}
{P. J.} Flory,
\newblock {\em J. Chem. Phys.}, {\bf 17}:303, 1949.

\bibitem{5}
{S. F.} Edwards,
\newblock {\em Proc. Phys. Soc.}, {\bf 88}:265, 1966;
\newblock {\em J. Phys.}, {\bf A 8}:1670, 1975.

\bibitem{6}
{P.-G} de~Gennes,
\newblock {\em J. Polym. Sci.}, {\bf 61}:313-315, 1977.

\bibitem{7}
J. Des Cloizeaux and G. Janninck,{\em Polymer in Solutions},
\newblock Oxford Univ. Press, New York, 1990.


\bibitem{8}
J.F. Joanny, P. Grant, L. Turkevich
and P. Pincus, \newblock {\em J. Phys. France}, {\bf 45}:151, 1984.

\bibitem{9}
{E.} Rapha{{\"e}}l, {G. H.} Fredrickson and {P. A.} Pincus,
\newblock {\em J. Phys. II France}, {\bf 2}:1811--1823, 1992.

\bibitem{JFFB}
{J.-F} Joanny and {F.} Brochard,
\newblock {\em J. Physique}, {\bf 42}:1145--1150, 1981.


\bibitem{10}
{E.} Helfand and {Y.} Tagami,
\newblock {\em J. Chem. Phys.}, {\bf 56}:3592, 1971.

\bibitem{11}
{P.-G} de~Gennes,
\newblock {\em Isr. J. Chem.}, {\bf 35}:33--35, 1995.


\bibitem{12}
{D. R. M.} Williams,
\newblock {\em Journal de Physique II}, {\bf 3}:1313--1318, 1993.


\bibitem{13}
{F. S.} Bates and {G. D.} Wignall,
\newblock {\em Phys. Rev. Lett.}, {\bf 57}:1429, 1991.

\bibitem{PGGCOLL}
{P.-G} de~Gennes,
\newblock {\em J. Physique Lett.}, {\bf 46}:L639--L642, 1985.

\bibitem{GROSBERG}
{A. Y.} Grosberg and {A. R.} Khokhlov,
\newblock {Giant molecules}, Academic Press, 1997.

\bibitem{KNOT}
{P.-G} de~Gennes,
\newblock {\em Macromol.}, {\bf 17}:703--704, 1984.





%\bibitem{Aczp}
%{S. F.} Edwards,
%\newblock {\em J. Phys.}, {\bf A 8}:1670, 1975.

%\bibitem{Bq}
%{P. J.} Flory, {\em Statistical Mechanics of Chain Molecules},
%\newblock Interscience Publishers, New York, 1969.

%\bibitem{Audp}
%{A.} Silberberg,
%\newblock {\em J. Coll. Interf. Sci.}, {\bf 901}:86--91, 1982.

%\bibitem{Uu}
%{P.-G.} {de~Gennes},
%\newblock Unpublished work, 1996.

%\bibitem{Aczt}
%{P. J.} Flory,
%\newblock {\em J. Chem. Phys.}, {\bf 10}:51, 1942.

%\bibitem{Aczq}
%{M. L.} Huggins,
%\newblock {\em J. Phys. Chem.}, {\bf 46}:151, 1942.

%\bibitem{Aczc}
%{M. L.} Huggins,
%\newblock {\em J. Am. Chem. Soc.}, {\bf 64}:1712, 1942.

\end{thebibliography}
%%%%%%%%%%%%%%%%%%%%%%%%%%%%%%%%%%%%%%%%%%%%%%%%%%%%%%%

%%%%%%%%%%%%%%%%%%%%%%%%%%%%%%%%%%%%
%%%%%%%%%%%%%%%%%%%%%%%%%%%%%%%%%%%%
\newpage
\begin{flushleft}
{\large {\bf Figure Captions}}.
\end{flushleft}
%%%%%%%%%%%%%%%%%%%%%%%%%%%%%%%%%%%%
{\bf Figure 1}. Flory free energy of a linear polymer molecule ($N$ monomers)
immersed in a melt of very long, chemically different chains, in the case of an
intermediate degree of incompatibility ($N^{-2/3} < \chi < N^{-1/2}$).
$R$ is the radius of gyration of the chain.
One minimum of the free energy corresponds to a
dense globule.
The other one corresponds to a Gaussian conformation.
%%%%%%%%%%%%%%%%%%%%%%%%%%%%%%%%%%%%
{\bf Figure 2}.
If the melt chains ($P$ monomers) are long, the $N$-chain displays
a hysteretic behavior when the degree of incompatibility is varied.
If the melt chain are shorter, their translational entropy enhances
interpenetration
with the $N$-chain : there is no more hysteresis when $P < N^{1/2}$.
For a simple solvent ($P = 1$), we recover the good, theta and
bad solvent regimes.

%%%%%%%%%%%%%%%%%%%%%%%%%%%%%%%%%%%%
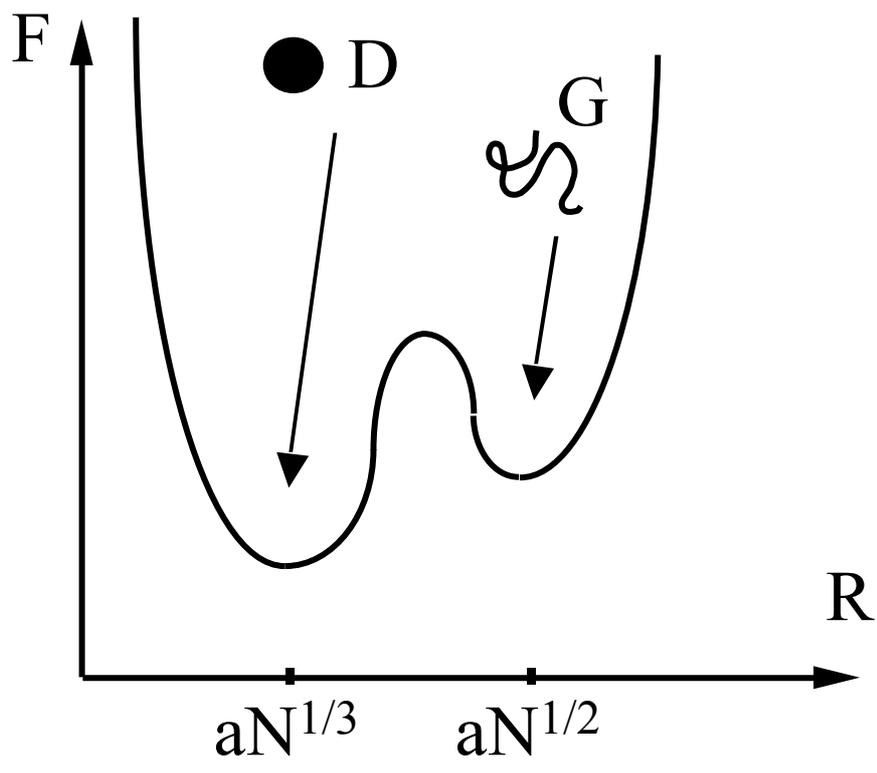
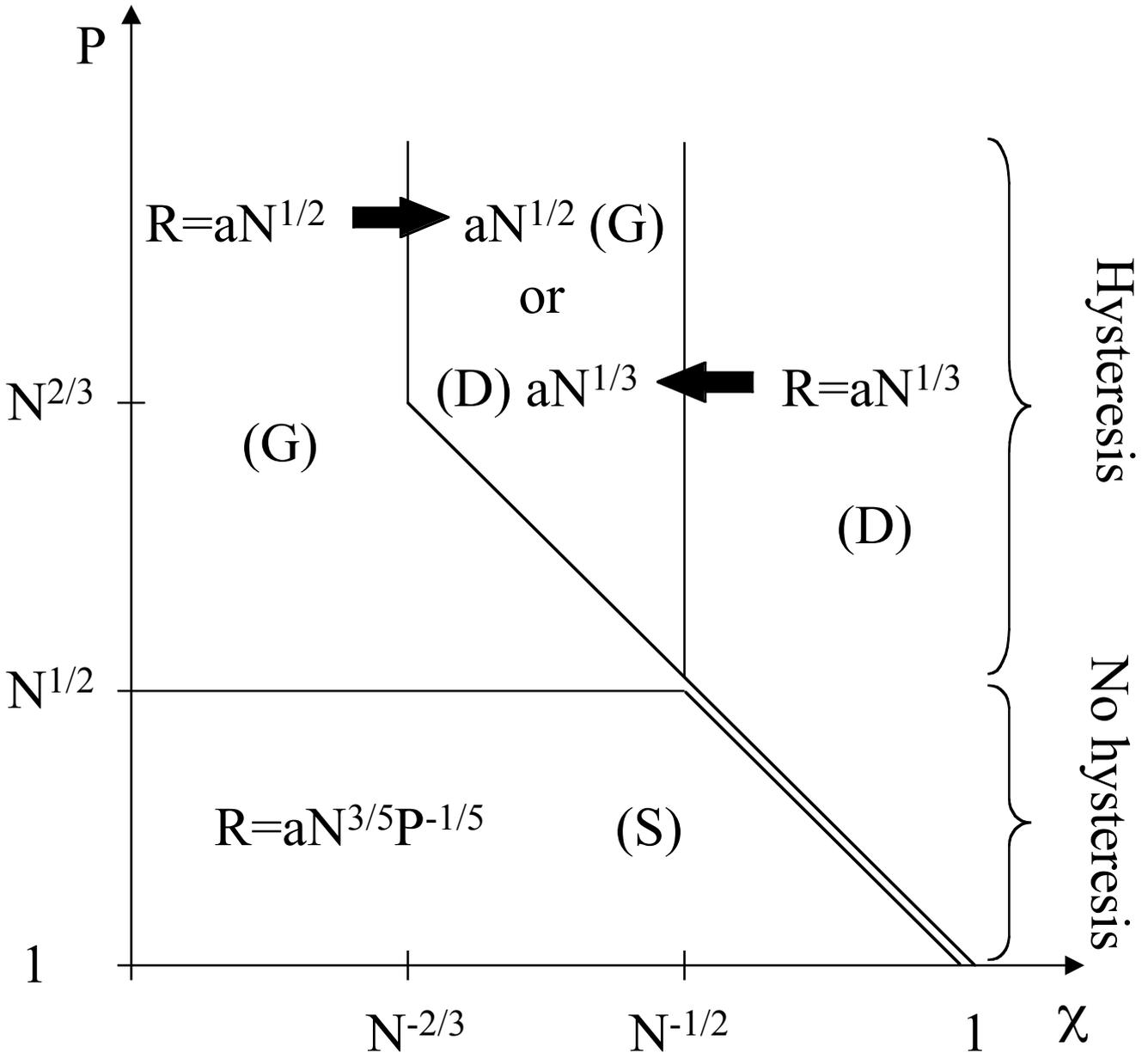
\end{document}